\input{aipcheck}

\documentclass[
    ,final            
  ]
  {aipproc}

\layoutstyle{8x11single}

\begin{document}

\title{Radon and material radiopurity assessment for the NEXT double beta decay experiment}


\makeatletter
\let\@fnsymbol=\@arabic
\makeatother

\author{S.~Cebri\'an}{
  address={Laboratorio de F\'isica Nuclear y Astropart\'iculas, Universidad de Zaragoza, C/ Pedro Cerbuna 12, 50009 Zaragoza, Spain}
,altaddress={Laboratorio Subterráneo de Canfranc, Paseo de los
Ayerbe s/n, 22880 Canfranc Estación, Huesca, Spain}} 
\author{J.~P\'erez}{
  address={Instituto de F\'isica Te\'orica, UAM/CSIC, Campus de Cantoblanco, 28049 Madrid, Spain} }
\author{I.~Bandac}{
  address={Laboratorio Subterráneo de Canfranc, Paseo de los Ayerbe s/n, 22880 Canfranc Estación, Huesca, Spain}}
\author{L.~Labarga}{
  address={Dpto. de F\'isica Te\'orica, Universidad Aut\'onoma de Madrid, Campus de Cantoblanco, 28049 Madrid, Spain}}
\author{V.~\'Alvarez}{
  address={Instituto de F\'isica Corpuscular, CSIC \& Universitat de Val\`encia, C/ Catedr\'atico Jos\'e Beltr\'an, 2, 46980 Paterna, Valencia,
Spain}}
\author{A.I.~Barrado}{
  address={Centro de Investigaciones Energ\'eticas, Medioambientales y Tecnol\'ogicas, Complutense 40, 28040 Madrid, Spain}}
\author{A.~Bettini}{
  address={Laboratorio Subterráneo de Canfranc, Paseo de los Ayerbe s/n, 22880 Canfranc Estación, Huesca, Spain},altaddress={Padua University and INFN Section, Dipartimento di Fisica G. Galilei, Via Marzolo 8, 35131 Padova, Italy}}
\author{F.I.G.M.~Borges}{
  address={Departamento de Fisica, Universidade de Coimbra, Rua Larga, 3004-516 Coimbra, Portugal}}
\author{M.~Camargo}{
  address={Centro de Investigaciones en Ciencias B\'asicas y Aplicadas, Universidad Antonio
  Nari\~no, Carretera 3 este No.\ 47A-15, Bogot\'a, Colombia}}
\author{S.~C\'arcel}{
  address={Instituto de F\'isica Corpuscular, CSIC \& Universitat de Val\`encia, C/ Catedr\'atico Jos\'e Beltr\'an, 2, 46980 Paterna, Valencia,
Spain}}
\author{A.~Cervera}{
  address={Instituto de F\'isica Corpuscular, CSIC \& Universitat de Val\`encia, C/ Catedr\'atico Jos\'e Beltr\'an, 2, 46980 Paterna, Valencia,
Spain}}
\author{C.A.N.~Conde}{
  address={Departamento de Fisica, Universidade de Coimbra, Rua Larga, 3004-516 Coimbra, Portugal}}
\author{E.~Conde}{
  address={Centro de Investigaciones Energ\'eticas, Medioambientales y Tecnol\'ogicas, Complutense 40, 28040 Madrid, Spain}}
\author{T.~Dafni}{
  address={Laboratorio de F\'isica Nuclear y Astropart\'iculas, Universidad de Zaragoza, C/ Pedro Cerbuna 12, 50009 Zaragoza, Spain}
,altaddress={Laboratorio Subterráneo de Canfranc, Paseo de los
Ayerbe s/n, 22880 Canfranc Estación, Huesca, Spain}}
\author{J.~D\'iaz}{
  address={Instituto de F\'isica Corpuscular, CSIC \& Universitat de Val\`encia, C/ Catedr\'atico Jos\'e Beltr\'an, 2, 46980 Paterna, Valencia,
Spain}}
\author{R.~Esteve}{
  address={Instituto de Instrumentaci\'on para Imagen Molecular, Universitat Polit\`ecnica de Val\`encia, Camino de Vera, s/n, Edificio 8B, 46022 Valencia, Spain}}
\author{L.M.P.~Fernandes}{
  address={Departamento de Fisica, Universidade de Coimbra, Rua Larga, 3004-516 Coimbra, Portugal}}
\author{M.~Fern\'andez}{
  address={Centro de Investigaciones Energ\'eticas, Medioambientales y Tecnol\'ogicas, Complutense 40, 28040 Madrid, Spain}}
\author{P.~Ferrario}{
  address={Instituto de F\'isica Corpuscular, CSIC \& Universitat de Val\`encia, C/ Catedr\'atico Jos\'e Beltr\'an, 2, 46980 Paterna, Valencia,
Spain}}
\author{E.D.C.~Freitas}{
  address={Institute of Nanostructures, Nanomodelling and Nanofabrication, Universidade de Aveiro, Campus de Santiago, 3810-193 Aveiro, Portugal}}
\author{L.M.P.~Fernandes}{
  address={Departamento de Fisica, Universidade de Coimbra, Rua Larga, 3004-516 Coimbra, Portugal}}
\author{V.M.~Gehman}{
  address={Lawrence Berkeley National Laboratory, 1 Cyclotron Road, Berkeley, California 94720, USA}}
\author{A.~Goldschmidt}{
  address={Lawrence Berkeley National Laboratory, 1 Cyclotron Road, Berkeley, California 94720, USA}}
\author{J.J.~G\'omez-Cadenas}{
  address={Instituto de F\'isica Corpuscular, CSIC \& Universitat de Val\`encia, C/ Catedr\'atico Jos\'e Beltr\'an, 2, 46980 Paterna, Valencia,
Spain}} 
\author{D.~Gonz\'alez-D\'iaz}{
  address={Laboratorio de F\'isica Nuclear y Astropart\'iculas, Universidad de Zaragoza, C/ Pedro Cerbuna 12, 50009 Zaragoza, Spain}
,altaddress={Laboratorio Subterráneo de Canfranc, Paseo de los
Ayerbe s/n, 22880 Canfranc Estación, Huesca, Spain}}
\author{R.M.~Guti\'errez}{
  address={Centro de Investigaciones en Ciencias B\'asicas y Aplicadas, Universidad Antonio
  Nari\~no, Carretera 3 este No.\ 47A-15, Bogot\'a, Colombia}}
\author{J.~Hauptman}{
  address={Department of Physics and Astronomy, Iowa State University, 12 Physics Hall, Ames, Iowa 50011-3160, USA}}
\author{J.A.~Hernando Morata}{
  address={Instituto Gallego de F\'isica de Altas Energ\'ias, Univ.\ de Santiago de
  Compostela, Campus sur, R\'ua Xos\'e Mar\'ia Su\'arez N\'u\~nez, s/n, 15782 Santiago de Compostela, Spain}}
\author{D.C.~Herrera}{
  address={Laboratorio de F\'isica Nuclear y Astropart\'iculas, Universidad de Zaragoza, C/ Pedro Cerbuna 12, 50009 Zaragoza, Spain}
,altaddress={Laboratorio Subterráneo de Canfranc, Paseo de los
Ayerbe s/n, 22880 Canfranc Estación, Huesca, Spain}}
\author{I.G.~Irastorza}{
  address={Laboratorio de F\'isica Nuclear y Astropart\'iculas, Universidad de Zaragoza, C/ Pedro Cerbuna 12, 50009 Zaragoza, Spain}
,altaddress={Laboratorio Subterráneo de Canfranc, Paseo de los
Ayerbe s/n, 22880 Canfranc Estación, Huesca, Spain}}
\author{A.~Laing}{
  address={Instituto de F\'isica Corpuscular, CSIC \& Universitat de Val\`encia, C/ Catedr\'atico Jos\'e Beltr\'an, 2, 46980 Paterna, Valencia,
Spain}}
\author{I.~Liubarsky}{
  address={Instituto de F\'isica Corpuscular, CSIC \& Universitat de Val\`encia, C/ Catedr\'atico Jos\'e Beltr\'an, 2, 46980 Paterna, Valencia,
Spain}}
\author{N.~L\'opez-March}{
  address={Instituto de F\'isica Corpuscular, CSIC \& Universitat de Val\`encia, C/ Catedr\'atico Jos\'e Beltr\'an, 2, 46980 Paterna, Valencia,
Spain}}
\author{D.~Lorca}{
  address={Instituto de F\'isica Corpuscular, CSIC \& Universitat de Val\`encia, C/ Catedr\'atico Jos\'e Beltr\'an, 2, 46980 Paterna, Valencia,
Spain}}
\author{M.~Losada}{
  address={Centro de Investigaciones en Ciencias B\'asicas y Aplicadas, Universidad Antonio
  Nari\~no, Carretera 3 este No.\ 47A-15, Bogot\'a, Colombia}}
\author{G.~Luz\'on}{
  address={Laboratorio de F\'isica Nuclear y Astropart\'iculas, Universidad de Zaragoza, C/ Pedro Cerbuna 12, 50009 Zaragoza, Spain}
,altaddress={Laboratorio Subterráneo de Canfranc, Paseo de los
Ayerbe s/n, 22880 Canfranc Estación, Huesca, Spain}}
\author{A.~Mar\'i}{
  address={Instituto de Instrumentaci\'on para Imagen Molecular, Universitat Polit\`ecnica de Val\`encia, Camino de Vera, s/n, Edificio 8B, 46022 Valencia, Spain}}
\author{J.~Mart\'in-Albo}{
  address={Instituto de F\'isica Corpuscular, CSIC \& Universitat de Val\`encia, C/ Catedr\'atico Jos\'e Beltr\'an, 2, 46980 Paterna, Valencia,
Spain}}
\author{A.~Mart\'inez}{
  address={Instituto de F\'isica Corpuscular, CSIC \& Universitat de Val\`encia, C/ Catedr\'atico Jos\'e Beltr\'an, 2, 46980 Paterna, Valencia,
Spain}}
\author{G.~Mart\'inez-Lema}{
  address={Instituto Gallego de F\'isica de Altas Energ\'ias, Univ.\ de Santiago de
  Compostela, Campus sur, R\'ua Xos\'e Mar\'ia Su\'arez N\'u\~nez, s/n, 15782 Santiago de Compostela, Spain}}
\author{T.~Miller}{
  address={Lawrence Berkeley National Laboratory, 1 Cyclotron Road, Berkeley, California 94720, USA}}
\author{F.~Monrabal}{
  address={Instituto de F\'isica Corpuscular, CSIC \& Universitat de Val\`encia, C/ Catedr\'atico Jos\'e Beltr\'an, 2, 46980 Paterna, Valencia,
Spain}}
\author{M.~Monserrate}{
  address={Instituto de F\'isica Corpuscular, CSIC \& Universitat de Val\`encia, C/ Catedr\'atico Jos\'e Beltr\'an, 2, 46980 Paterna, Valencia,
Spain}}
\author{C.M.B.~Monteiro}{
  address={Departamento de Fisica, Universidade de Coimbra, Rua Larga, 3004-516 Coimbra, Portugal}}
\author{F.J.~Mora}{
  address={Instituto de Instrumentaci\'on para Imagen Molecular, Universitat Polit\`ecnica de Val\`encia, Camino de Vera, s/n, Edificio 8B, 46022 Valencia, Spain}}
\author{L.M. Moutinho}{
  address={Institute of Nanostructures, Nanomodelling and Nanofabrication, Universidade de Aveiro, Campus de Santiago, 3810-193 Aveiro, Portugal}}
\author{J.~Mu\~noz Vidal}{
  address={Instituto de F\'isica Corpuscular, CSIC \& Universitat de Val\`encia, C/ Catedr\'atico Jos\'e Beltr\'an, 2, 46980 Paterna, Valencia,
Spain}}
\author{M.~Nebot-Guinot}{
  address={Instituto de F\'isica Corpuscular, CSIC \& Universitat de Val\`encia, C/ Catedr\'atico Jos\'e Beltr\'an, 2, 46980 Paterna, Valencia,
Spain}}
\author{D.~Nygren}{
  address={Lawrence Berkeley National Laboratory, 1 Cyclotron Road, Berkeley, California 94720, USA}}
\author{C.A.B.~Oliveira}{
  address={Lawrence Berkeley National Laboratory, 1 Cyclotron Road, Berkeley, California 94720, USA}}
\author{A. Ortiz de Sol\'orzano}{
  address={Laboratorio de F\'isica Nuclear y Astropart\'iculas, Universidad de Zaragoza, C/ Pedro Cerbuna 12, 50009 Zaragoza, Spain}
,altaddress={Laboratorio Subterráneo de Canfranc, Paseo de los
Ayerbe s/n, 22880 Canfranc Estación, Huesca, Spain}}
\author{J.L.~P\'erez Aparicio}{
  address={Dpto.\ de Mec\'anica de Medios Continuos y Teor\'ia de Estructuras, Univ.\ Polit\`ecnica de Val\`encia, Camino de Vera, s/n, 46071 Valencia, Spain}}
\author{M.~Querol}{
  address={Instituto de F\'isica Corpuscular, CSIC \& Universitat de Val\`encia, C/ Catedr\'atico Jos\'e Beltr\'an, 2, 46980 Paterna, Valencia,
Spain}}
\author{J.~Renner}{
  address={Lawrence Berkeley National Laboratory, 1 Cyclotron Road, Berkeley, California 94720, USA}}
\author{L.~Ripoll}{
  address={Escola Polit\`ecnica Superior, Universitat de Girona, Av.~Montilivi, s/n, 17071 Girona, Spain}}
\author{J.~Rodr\'iguez}{
  address={Instituto de F\'isica Corpuscular, CSIC \& Universitat de Val\`encia, C/ Catedr\'atico Jos\'e Beltr\'an, 2, 46980 Paterna, Valencia,
Spain}}
\author{F.P.~Santos}{
  address={Departamento de Fisica, Universidade de Coimbra, Rua Larga, 3004-516 Coimbra, Portugal}}
\author{J.M.F.~dos Santos}{
  address={Departamento de Fisica, Universidade de Coimbra, Rua Larga, 3004-516 Coimbra, Portugal}}
\author{L.~Serra}{
  address={Instituto de F\'isica Corpuscular, CSIC \& Universitat de Val\`encia, C/ Catedr\'atico Jos\'e Beltr\'an, 2, 46980 Paterna, Valencia,
Spain}}
\author{D.~Shuman}{
  address={Lawrence Berkeley National Laboratory, 1 Cyclotron Road, Berkeley, California 94720, USA}}
\author{A. Sim\'on}{
  address={Instituto de F\'isica Corpuscular, CSIC \& Universitat de Val\`encia, C/ Catedr\'atico Jos\'e Beltr\'an, 2, 46980 Paterna, Valencia,
Spain}}
\author{C.~Sofka}{
  address={Department of Physics and Astronomy, Texas A\&M University, College Station, Texas 77843-4242, USA}}
\author{M.~Sorel}{
  address={Instituto de F\'isica Corpuscular, CSIC \& Universitat de Val\`encia, C/ Catedr\'atico Jos\'e Beltr\'an, 2, 46980 Paterna, Valencia,
Spain}}
\author{J.F.~Toledo}{
  address={Instituto de Instrumentaci\'on para Imagen Molecular, Universitat Polit\`ecnica de Val\`encia, Camino de Vera, s/n, Edificio 8B, 46022 Valencia, Spain}}
\author{J.~Torrent}{
  address={Escola Polit\`ecnica Superior, Universitat de Girona, Av.~Montilivi, s/n, 17071 Girona, Spain}}
\author{Z.~Tsamalaidze}{
  address={Joint Institute for Nuclear Research, Joliot-Curie 6, 141980 Dubna, Russia}}
\author{J.F.C.A.~Veloso}{
  address={Institute of Nanostructures, Nanomodelling and Nanofabrication, Universidade de Aveiro, Campus de Santiago, 3810-193 Aveiro, Portugal}}
\author{J.A.~Villar}{
  address={Laboratorio de F\'isica Nuclear y Astropart\'iculas, Universidad de Zaragoza, C/ Pedro Cerbuna 12, 50009 Zaragoza, Spain}
,altaddress={Laboratorio Subterráneo de Canfranc, Paseo de los
Ayerbe s/n, 22880 Canfranc Estación, Huesca, Spain}}
\author{R.C.~Webb}{
  address={Department of Physics and Astronomy, Texas A\&M University, College Station, Texas 77843-4242, USA}}
\author{J.T.~White}{
  address={Department of Physics and Astronomy, Texas A\&M University, College Station, Texas 77843-4242, USA}}
\author{N.~Yahlali}{
  address={Instituto de F\'isica Corpuscular, CSIC \& Universitat de Val\`encia, C/ Catedr\'atico Jos\'e Beltr\'an, 2, 46980 Paterna, Valencia,
Spain}}

\begin{abstract}
The ''Neutrino Experiment with a Xenon TPC'' (NEXT), intended to
investigate the neutrinoless double beta decay using a high-pressure
xenon gas TPC filled with Xe enriched in $^{136}$Xe at the Canfranc
Underground Laboratory in Spain, requires ultra-low background
conditions demanding an exhaustive control of material radiopurity
and environmental radon levels. An extensive material screening
process is underway for several years based mainly on gamma-ray
spectroscopy using ultra-low background germanium detectors in
Canfranc but also on mass spectrometry techniques like GDMS and
ICPMS. Components from shielding, pressure vessel,
electroluminescence and high voltage elements and energy and
tracking readout planes have been analyzed, helping in the final
design of the experiment and in the construction of the background
model. The latest measurements carried out will be presented and the
implication on NEXT of their results will be discussed. The
commissioning of the NEW detector, as a first step towards NEXT, has
started in Canfranc; in-situ measurements of airborne radon levels
were taken there to optimize the system for radon mitigation and
will be shown too.
\end{abstract}

\maketitle


\section{Introduction}

The observation of neutrinoless double beta decay would be
outstanding for characterizing neutrino properties \cite{dbdrefs}.
The NEXT experiment (``\underline{N}eutrino \underline{E}xperiment
with a \underline{X}enon \underline{T}ime-Projection Chamber'') aims
to search for such a decay in $^{136}$Xe at the Laboratorio
Subterráneo de Canfranc (LSC), located at the Spanish Pyrenees, with
a source mass of $\sim$100 kg (NEXT-100 phase). The challenge is to
combine, while keeping the detector$=$source approach, the
measurement of the topological signature of the event (in order to
discriminate the signal from background) with the energy resolution
optimization (to single out the peak at the sum energy of the two
emitted electrons). The NEXT detector will be a high pressure
gaseous xenon Time-Projection Chamber (TPC) with proportional
electroluminescent (EL) amplification \cite{next}. There will be
separate energy and tracking readout planes, located at opposite
sides of the pressure vessel, using different sensors:
photomultiplier tubes (PMTs) for calorimetry (and for fixing the
start of the event) and silicon photomultipliers (SiPMs) for
tracking. While work on prototypes is still ongoing \cite{protos},
the installation of shielding and ancillary system started at LSC in
2013. Underground commissioning of the NEW detector began at the end
of 2014 and first data are expected along 2015. The NEW (NEXT-WHITE)
apparatus\footnote{The name honours the memory of the late Professor
James White, key scientist of the NEXT project.} is the first phase
of the NEXT detector to operate underground; it is a downscale 1:2
in size (1:8 in mass) of NEXT-100.

The goal of NEXT is to explore electron neutrino effective Majorana
masses below 100 meV for a total exposure of 500 kg$\cdot$year. To
reach this sensitivity, there are two basic requirements: 1) An
energy resolution of at most 1\% FWHM at the transition energy
(Q$_{\beta\beta}=$2.458 MeV), which is reachable with EL
amplification according to the results of prototypes. 2) A
background level below $8\times10^{-4}$ counts keV$^{-1}$ kg$^{-1}$
y$^{-1}$ in the energy region of interest, achievable thanks to
passive shieldings, pattern recognition techniques and a thorough
material radiopurity control. Materials to be used in the whole
experimental set-up have been screened and first results presented
in \cite{jinstrp,aiprp,javi,icheprp,trackingrp}. Here, new results
are shown and discussed together with the first direct
quantification of radon levels at the NEXT site in Canfranc.

\section{Material radioassay}

The material screening program of the NEXT experiment is mainly
based on germanium $\gamma$-ray spectrometry using ultra-low
background detectors operated deep underground, at a depth of 2450
m.w.e., from the Radiopurity Service of LSC; being a non-destructive
technique, the actual components to be used in the experiment can be
analyzed. Detectors are p-type close-end coaxial 2.2-kg High Purity
germanium detectors, from Canberra France.
For the measurements presented here, the shielding consisted of 5 or
10 cm of copper in the inner part surrounded by 20 cm of low
activity lead, with boil-off nitrogen flush to avoid airborne radon
intrusion. Detection efficiency is estimated for each sample by
GEANT4 simulation. To complement germanium spectrometry results,
measurements based on Glow Discharge Mass Spectrometry (GDMS) and
Inductively Coupled Plasma Mass Spectrometry (ICPMS) have been also
carried out. GDMS is performed by Evans Analytical Group in France,
providing concentrations of U, Th and K. ICPMS measurements were
made at CIEMAT (Unidad de Espectrometria de Masas) in Spain.

Materials analyzed deal with the shielding, pressure vessel, field
cage and EL components and the energy and tracking readout planes.
Results obtained after those already presented in
\cite{jinstrp}-\cite{trackingrp} are summarized in table~\ref{rpm}
and described in the following; for germanium measurements, reported
errors correspond to 1$\sigma$ uncertainties and upper limits are
given at 95\% C.L.. Uncertainties for GDMS results are typically of
20\%.

The passive shielding of NEXT consists of a 20-cm-thick lead castle
together with an additional 12-cm-thick copper layer placed inside
the vessel. Lead and copper from different suppliers were studied
\cite{jinstrp,aiprp,icheprp}. The selected options were CuA1 (or
ETP) copper from Lugand Aciers and refurbished lead from the OPERA
experiment, following the analysis by GDMS and using Ge detectors
\cite{icheprp}. After the melting of the OPERA lead sheets for brick
production by Tecnibusa company a new screening of the material was
performed (row \#1 of table~\ref{rpm}); no contamination was
introduced in the process and the nominal value of 80 Bq/kg for
$^{210}$Pb activity \cite{opera} was verified by measuring
(76.4$\pm$8.6)~Bq/kg using the 46.5~keV line. Different types of
steel used in the lead castle structure have been analyzed. After
screening samples of the 316Ti stainless steel from Nironit used for
walls \cite{aiprp}, structural steel S-275 was also measured; three
samples were taken into consideration, made of the raw steel, steel
after coating with antioxidant primer and after additional painting
(rows \#2-4 of table~\ref{rpm}). The primer seems to be responsible
for the high activities observed. Materials intended to be used for
filling empty space in the shielding were also screened: lead wool
from Tecnibusa and foam joints made of EPDM sponge from Moss Express
and Artein Gaskets (rows \#5-7 of table~\ref{rpm}). It has been
checked that the activity measured for these components should not
have an impact on the background model of the experiment.

For the pressure vessel of NEXT-100, which must hold 15 b, several
samples of 316Ti stainless steel from Nironit were initially
screened with germanium detectors \cite{jinstrp} for body, end-caps
and flanges and complementary results were obtained afterwards from
GDMS analysis \cite{icheprp}. The TIG-MIG welding was also analyzed
\cite{aiprp} and recently a sample of glue intended to be used on
the tests of the vessel welding was screened (row \#8 of table
\ref{rpm}).

The field cage is made using copper rings connected to resistors and
High Density polyethylene as insulator; wire meshes separate the
different field regions and reflector panels made of PTFE and coated
with a wavelength shifter (TPB) improve the light collection. Many
types of plastics (peek, semitron, kynar, tefzel and poliethylene
from different suppliers) were screened and HD polyethylene was
analyzed also by ICPMS \cite{jinstrp,aiprp,icheprp}. ETP copper for
field cage, in rod and sheet, was measured using GDMS \cite{icheprp}
and now results for HV chip resistors from Ohmcraft can be presented
too (row \#9 of table~\ref{rpm}); activities higher than expected
were found and more radiopure resistors will be searched for
NEXT-100.

The energy readout plane of NEXT-100 will include 60 Hamamatsu
R11410-10 PMTs placed behind the cathode to detect EL light and
primary scintillation. Each PMT is sealed into individual, pressure
resistant copper cans, coupled to the sensitive volume through a
sapphire window coated with TPB. Shappire windows, copper for cans
and plates and several components for PMT bases (capacitors,
resistors, pin receptacles and thermal epoxy) were analyzed
\cite{aiprp,icheprp}. Other items have been taken into consideration
using Ge detectors or by GDMS analysis and results are presented in
table~\ref{rpm}: brazing paste (made of 72\% Ag and 28\% Cu, row
\#10), M4 stainless steel vented screws and brass bolts (for PMT
cans, rows \#11-12), the optical coupling gel (SmartGel NyoGel
OCK-451, row \#13), other epoxy (Araldite 2011, mixing resin and
hardener, row \#14) and kapton-copper cable (used at bases, row
\#15). U and Th concentration of brass bolts are about two orders of
magnitude lower than those of stainless steel screws and therefore
brass units have been selected. For Araldite 2011 epoxy, the
obtained results are more stringent than those available at
\cite{busto}. Presence of $^{108m}$Ag was also identified in the
kapton-copper cable. The available PMTs of the selected model for
NEXT-100 are being screened in 3-unit groups using the same Ge
detector; since all of them show equivalent activity, a joint
analysis of fifteen available runs (corresponding to 45 PMTs) has
been carried out to derive activities per PMT (row \#16 of
table~\ref{rpm}). The results are compatible with those presented by
XENON \cite{xenon} and PandaX collaborations \cite{pandax}; the same
PMT model has been also analyzed in \cite{lux,arisaka}. The modified
version R11410-21 of the Hamamatsu PMTs presents a $^{60}$Co
activity reduced by a factor $\sim$4-5, according to XENON1T results
\cite{xenon1t}.


The tracking readout plane of NEXT-100 will consist of an array of
110 boards (named ``Dice Boards'', DB) placed behind the EL region.
Each DB contains 8$\times$8 SiPM sensors with a pitch of $\sim$1 cm
and is coated with TPB. The design of a radiopure tracking plane, in
direct contact with the gas detector medium, was a challenge since
the needed components have typically activities too large for
experiments requiring ultra-low background conditions; results of
the radiopurity assessment for the tracking plane are presented in
\cite{trackingrp}. The substrate of the DBs is made of kapton and
copper from Flexible Circuits Inc., showing better radiopurity than
cuflon boards. SiPMs from SensL company (MLP, Moulded Lead-frame
Package plastic SMT elements, MicroFC-10035-SMT-GP) were chosen
after successful radioassay. Different electronic components
(capacitors, connectors, solder paste, NTC temperature sensors and
blue LEDs) were also screened with germanium detectors. New results
have been obtained for the board-to-board connectors (HIROSE
FX11LA-140S-SV, receptacles and HIROSE FX11LA-140P-SV, headers, row
\#17 of table \ref{rpm}) finding similar values to those obtained
for other connectors also made of LCP \cite{trackingrp}; these
connectors will be placed behind a copper shield. In an attempt to
identify the origin of the measured activity in the kapton-Cu DBs
(even if acceptable~according to the background model of NEXT-100)
the adhesive films used in these DBs were separately screened (row
\#18 of table~\ref{rpm}); the observed emissions cannot explain the
activity of the boards.

\begin{table}
\begin{tabular}{p{0.1cm}p{2.2cm}p{2.1cm}p{1.2cm}p{1cm}p{1.7cm}p{1.4cm}p{1.4cm}p{1.4cm}p{1cm}p{1.7cm}p{1cm}p{0.8cm}}
\hline &  Material    &Supplier   &Technique& units&  $^{238}$U &
$^{226}$Ra &    $^{232}$Th & $^{228}$Th &   $^{235}$U & $^{40}$K &
$^{60}$Co & $^{137}$Cs \\ \hline
1&  Lead & Britannia&  Ge& mBq/kg& $<$126&   $<$2.8    & &  $<$3.2    &&   $<$6.9  &$<$0.3&  \\
2& S-275 steel & Proycon & Ge & mBq/kg& 32$\pm$9 & 1.2$\pm$0.1& 1.9$\pm$0.2& 4.7$\pm$0.3 & & 3.2$\pm$0.7 & 1.8$\pm$0.1 & $<$0.2\\
3& Steel+primer & Proycon & Ge & mBq/kg& (1.1$\pm$0.3)$\times 10^{3}$ &444$\pm$21 & 125$\pm$9 & 106$\pm$6 & & (1.6$\pm$0.2)$\times 10^{3}$ &  94$\pm$7&$<$3.9 \\
4& Steel+primer & Proycon & Ge & mBq/kg& (0.8$\pm$0.2)$\times 10^{3}$& 437$\pm$20 & 76$\pm$5 & 58$\pm$3 & & (1.2$\pm$0.1)$\times 10^{3}$ & 2.2$\pm$0.3& $<$1.4\\
& +painting & & & & & & & & & & \\

5 & Lead wool & Tecnibusa & Ge & mBq/kg& $<$368 & $<$12 & & $<$15 &&
36.9$\pm$6.5 & $<$1.1 & \\
6 & EPDM foam & Moss Express & Ge &mBq/m & $<$437& 33.0$\pm$1.7 &106$\pm$7 & 95.5$\pm$5.2 & &758$\pm$78 &  $<$1.4 & $<$1.3 \\
7 & EPDM foam & Artein Gaskets & Ge & mBq/m &  $<$215  & 4.3$\pm$0.4
& 5.1$\pm$1.6 & $<$5.4 & & 11.2$\pm$2.9 & $<$0.6 & $<$0.6 \\ \hline

8 & Glue & Ceys & Ge & mBq/kg & $<$3.2$\times 10^{3}$ & $<$18 &
$<$75 & $<$31 & $<$13 & (33.0$\pm$3.3)$\times 10^{3}$ & $<$12 &
$<$10
\\ \hline

9 & Resistors & Ohmcraft & Ge & $\mu$Bq/pc & (0.56 $\pm$0.15)$\times
10^{3}$ & 217$\pm$10 & 44$\pm$4 & 36$\pm$3 & & 95$\pm$13 & $<$2 &
$<$2 \\ \hline

10 & Brazing paste & &GDMS & $\mu$Bq/kg & 55$\pm$10 && 49$\pm$4 &&& $<$31 & & \\
11 & Brass bolts & &GDMS & $\mu$Bq/kg & 8.9$\pm$0.7 && 6.9$\pm$0.2 & & &  $<$31 & & \\
12 & SS screws & &GDMS & mBq/kg & 3.25$\pm$0.25 & & 0.57$\pm$0.08 & && $<$0.19 & & \\
13 & Optical gel & Nye Lubricants & Ge & mBq/kg & $<$1.7$\times10^{3}$ & $<$22 & $<$49 & $<$18 & $<$16 & $<$173 & $<$4.5 & $<$5.8 \\
14 & Epoxy & Araldite & Ge & mBq/kg & $<$182 & $<$1.4 & $<$3.7 &
$<$2.5 & $<$0.8 & 15.0$\pm$2.4 & $<$0.4 &  $<$0.4 \\
15 & Kapton-Cu cable & & Ge & mBq/kg & $<$1.1$\times10^{3}$ & 46.8$\pm$3.3 & $<$40 & $<$32 & & 166$\pm$27 & $<$5.2 & $<$4.4 \\
16 & PMTs  & Hamamatsu & Ge & mBq/pc & $<$67 & $<$0.94 & $<$2.2 &
0.56$\pm$0.14 & 0.58$\pm$0.13  &  11.8$\pm$1.7 & 3.73$\pm$0.27 &
$<$0.3 \\
& (R11410-10) & & & & & & & & & & \\ \hline
17 & Connectors & Hirose & Ge & mBq/pc & 6.4$\pm$1.9 & 2.8$\pm$0.1 & 5.6$\pm$0.3  & 5.9$\pm$0.3 & & 3.4$\pm$0.4 &  $<$0.03 & $<$0.04 \\
18 & Adhesive films & Flexible Circuits & Ge & mBq/kg &
(1.8$\pm$0.6)$\times10^{3}$ & $<$19 & $<$50 &  $<$34 &  16.8$\pm$3.0
& $<$107 & $<$4.8 & $<$4.6 \\ \hline
\end{tabular}
\caption{Activities recently measured in relevant materials for
NEXT. GDMS results were derived from U, Th and K concentrations.
Germanium $\gamma$-ray spectrometry results reported for $^{238}$U
and $^{232}$Th correspond to the upper part of the chains (derived
from $^{234m}$Pa and $^{228}$Ac emissions) and those of $^{226}$Ra
and $^{228}$Th give activities of the lower parts.} \label{rpm}
\end{table}

\section{Control of radon}

A series of measurements of radon activity levels and gamma
background counting rates at the NEXT site in Canfranc has been
performed to quantify and understand the evolution of these
background components, with the aim to efficiently design their
mitigation, profiting from the fact of having the external lead
shielding of NEXT assembled at LSC without the detector mounted
inside for several months. For first radon measurements an
AlphaGuard counter from LSC was used inside NEXT lead castle,
together with another one which takes data routinely at hall A,
allowing to monitor radon evolution. To quantify the gamma
background a simple set-up was prepared, consisting of a
$3''\times3''$ low background NaI(Tl) detector followed by a
Canberra 2005 preamplifier, a Tennelec linear amplifier and a
Canberra 8701 ADC, read by an Arduino board. In this way, energy of
events up to $\sim$3~MeV was continuously registered.

From October 2014 to February 2015, data were taken
simultaneously by radon and gamma detectors at the following
distinct conditions:
\begin{enumerate}
\item With the lead castle open.
\item After closing the lead castle.
\item After improving the conditions for closing the lead castle by blocking holes and using additional foam joints.
\item After performing a N$_{2}$ purge, injecting a high flux of N$_{2}$ gas (1800 l$/$h) for about seven hours. In this way, the total N$_{2}$ gas injected was twice the internal volume of the shielding.
\item With the injection of a low constant N$_{2}$ gas flux (180 l/h), set after performing a second N$_{2}$ purge equivalent to the first one.
\item After stopping the constant N$_{2}$ gas flux in situation 5.
\item With the injection of a high constant N$_{2}$ gas flux (900 l/h), set also after a new N$_{2}$ purge equivalent to the previous ones.
\end{enumerate}

Table \ref{situations} shows the total counting rate of the NaI(Tl)
detector obtained at all the conditions of measurement, during
stable~periods following the different actions performed. Radon mean
activity values, measured by the AlphaGuard detectors at these
conditions both at hall A and inside the lead NEXT castle, are given
in table~\ref{situations} too; the corresponding standard deviations
are indicated. By closing the castle, the gamma rate was reduced two
orders of magnitude and the radon activity more than a factor 2,
becoming insensitive to external fluctuations. A N$_{2}$ purge and
even a low constant N$_{2}$ gas flux produced only marginal
reductions. A high constant N$_{2}$ gas flux allowed a more
significant reduction, of at least one order of magnitude for the
radon activity inside the castle; indeed, the sensitivity limit of
the AlphaGuard detector was reached in that situation. A more
sensitive radon detector would be necessary to quantify the actual
reduction thanks to this high constant N$_{2}$ gas rate; data were
taken contemporaneously also using a custom-made radon detector with
micromegas readout and analysis of these data is underway.

\begin{table}
\caption{Summary of measurements taken at different conditions at
the NEXT site in Canfranc: total rate registered by the NaI(Tl)
detector and radon mean activities measured by the AlphaGuard
detectors both at hall A and inside the lead castle. Standard
deviations ($\sigma$) for radon activity are also shown.}
\begin{tabular}{|cl|c|cc|cc|}
\hline &Situation  &  NaI(Tl) detector rate & Rn activity & at Hall
A & Rn activity & at NEXT
castle \\
&&(Hz) & (Bq$/$m$^{3}$) & & (Bq$/$m$^{3}$) & \\
&&& mean & $\sigma$ & mean & $\sigma$ \\ \hline

1 & Open castle & 59.56$\pm$0.01&  & & &  \\

2& Closed castle &1.089$\pm$0.001 & 80 & 29 &79 & 28 \\

3& Better closed castle &0.694$\pm$0.002 & 74 & 26 & 30 & 11  \\

4& After N$_{2}$ purge &  0.658$\pm$0.001& 66 & 25 & 30 & 12   \\

5& N$_{2}$ purge+flux 180 l/h & 0.600$\pm$0.001 & 66 & 25 & 22 & 9 \\

6 & Without N$_{2}$ flux &0.638$\pm$0.001  &  59 & 18 & 26 & 10 \\

7 & N$_{2}$ purge+flux 900 l/h &0.350$\pm$0.002  & 73 & 26 & 4 &  3\\ \hline

\end{tabular}
\label{situations}
\end{table}

\section{Summary}
An extensive screening program for materials and components used in
the NEXT experiment is underway since 2011, based mainly on
ultra-low background gamma spectrometry using HPGe detectors at LSC
but also on complementary results from GDMS and ICPMS. Information
on radiopurity obtained has helped in the design of the set-up:
adequate materials for shieldings, vessel and field cage have been
identified and a thorough selection of in-vessel components for
energy and tracking planes has been performed too, checking
radiopurity of all photomultiplier units and choosing finally
kapton-copper boards and optimum SiPMs. Measured activity of the
relevant radioisotopes has been used also as input in the
construction of the background model of NEXT-100 \cite{ichep};
according to first results, the required background level seems to
be at reach.

On the other hand, airborne radon activity and gamma background
rates have been measured using different detectors at various
conditions inside the NEXT lead castle in Canfranc, before the
commissioning of the NEW detector there, in order to optimize the
radon suppression system.

\begin{theacknowledgments}
The NEXT Collaboration acknowledges funding support from the
following agencies and institutions: the European Research Council
under the Advanced Grant 339787-NEXT and the T-REX Starting Grant
ref. ERC-2009-StG-240054 of the IDEAS program of the 7th EU
Framework Program; the Spanish Ministerio de Economía y
Competitividad under grants CONSOLIDER-Ingenio 2010 CSD2008-0037
(CUP), FPA2009-13697-C04-04, and FIS2012-37947-C04; the Director,
Office of Science, Office of Basic Energy Sciences of the US DoE
under Contract no. DE-AC02-05CH11231; and the Portuguese FCT and
FEDER through the program COMPETE, Projects PTDC/FIS/103860/2008 and
PTDC/FIS/112272/2009. S. Cebrián acknowledges the support from the
University of Zaragoza (Convocatoria propia de proyectos de
investigación para Jóvenes Investigadores 2014). Special thanks are
due to LSC directorate and staff for their strong support for
performing the measurements at the LSC Radiopurity Service.
\end{theacknowledgments}

\bibliographystyle{aipproc}

\end{document}